\documentclass[aps,prd,print,groupedaddress,nofootinbib,eqsecnum,onecolumn]{revtex4}
\usepackage{eurosym}
\usepackage{amsmath}
\usepackage{amsthm}
\usepackage{amssymb}
\usepackage{slashed}
\usepackage{hyperref}
\usepackage{epstopdf}

\usepackage[english]{babel}
\usepackage[latin1]{inputenc}
\usepackage{amsmath, amsthm, amssymb}
\usepackage[pdftex]{graphicx}
\usepackage{amssymb}
\usepackage{latexsym}
\usepackage{amstext}
\usepackage{epstopdf}
\usepackage{floatflt}
\usepackage{hyperref}
\usepackage{enumitem}
\usepackage{multirow}
\usepackage{pdflscape}
\usepackage{makecell}
\usepackage{nopageno}

\numberwithin{equation}{section}
\begin{document}

\begin{titlepage}
\begin{center}

\vskip 3.0cm

{\bf \huge The Magic of Being Exceptional}

\vskip 3.0cm

{\bf \large Alessio Marrani${}^{1}$, Piero Truini${}^{2}$, and Michael Rios${}^{3}$ }

\vskip 40pt

{\it ${}^1$Museo Storico della Fisica e Centro Studi e Ricerche ``Enrico Fermi'',\\
Via Panisperna 89A, I-00184, Roma, Italy}\\\vskip 5pt

\vskip 40pt

{\it ${}^2$Dipartimento di Fisica and INFN, Universit\`{a} di Genova,\\
Via Dodecaneso 33, I-16146 Genova, Italy}\\ \vskip 5pt

\vskip 40pt

{\it ${}^3$Dyonica ICMQG,\\5151 State University Drive, Los Angeles, CA 90032, USA}\\

\vskip 30pt

\texttt{truini@ge.infn.it},
\texttt{jazzphyzz@gmail.com},
\texttt{mrios@dyonicatech.com}

\end{center}

\vskip 95pt

\begin{center} {\bf ABSTRACT}\\[3ex]\end{center}

Starting from the Jordan algebraic interpretation of the \textit{%
\textquotedblleft Magic Star" }embedding within the exceptional sequence of
simple Lie algebras, we exploit the so-called spin factor embedding of rank-$%
3$ Jordan algebras and its consequences on the Jordan algebraic Lie
symmetries, in order to provide another perspective on the origin of the
\textit{Exceptional Periodicity (EP)} and its \textit{\textquotedblleft
Magic Star"} structure. We also highlight some properties of the special
class of Vinberg's rank-$3$ (dubbed \textit{exceptional}) T-algebras,
appearing on the tips of the \textit{\textquotedblleft Magic Star"}
projection of \textit{EP}(-generalized, finite-dimensional, exceptional)
\textit{algebras}.

\vskip 65pt

\begin{center}
Presented by A.M. at the \textit{32nd International Colloquium on Group Theoretical Methods in Physics},\\Prague, July 9-13, 2018
\end{center}





%
\vfill

\end{titlepage}

\newpage \setcounter{page}{1} \numberwithin{equation}{section}

\section{The Magic Star and the Exceptional Sequence}

\begin{figure}[h]
\centering
\includegraphics[width=0.60\textwidth]{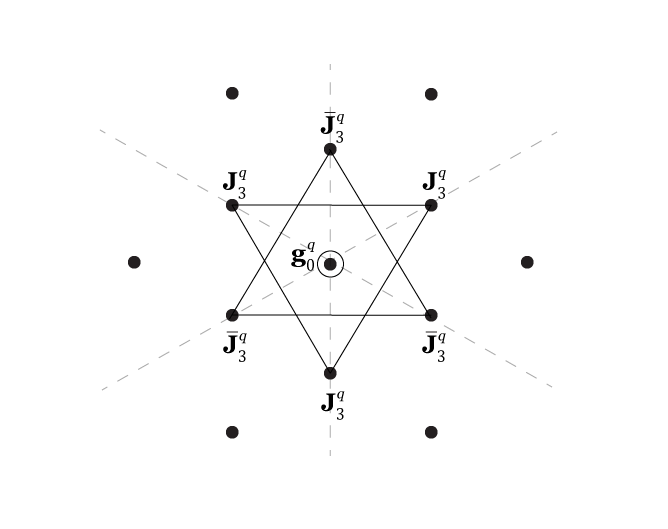}
\caption{The \textit{\textquotedblleft Magic Star"} of finite-dimensional
exceptional Lie algebras \protect\cite{mukai,pt1}. $\mathbf{J}_{3}^{q}$
denotes a simple Jordan algebra of rank-3, parametrized by $q=\dim _{\mathbb{%
R}}\mathbb{A}=1,2,4,8$ for $\mathbb{A}=\mathbb{R},\mathbb{C},\mathbb{H},%
\mathbb{O}$, corresponding to $\mathbf{f}_{4}$, $\mathbf{e}_{6}$, $\mathbf{e}%
_{7}$, $\mathbf{e}_{8}$, respectively. In the case of $\mathbf{g}_{2}$
(corresponding to $q=-2/3$), the Jordan algebra is trivially the identity
element : $\mathbf{J}_{3}^{-2/3}\equiv \mathbb{I}:=diag(1,1,1)$.}
\label{fig:MagicStar}
\end{figure}

Given a finite-dimensional exceptional Lie algebra, the so-called \textit{%
\textquotedblleft Magic Star" projection} (depicted\footnote{%
Note the slight change of notation with respect to \cite{3bis}, in which
also the equivalent notation $\mathfrak{C}$ (Cayley numbers) was used for
octonions $\mathbb{O}$.} in Fig. \ref{fig:MagicStar}) of the corresponding
root lattice onto a plane determined by an $\mathbf{a}_{\mathbf{2}}$ root
sub-lattice has been introduced by Mukai\footnote{%
Mukai used the name \textit{\textquotedblleft }$\mathbf{g}_{2}$ \textit{%
decomposition"} to denote the \textit{Magic Star} structure of exceptional
Lie algebras.} \cite{mukai}, and later investigated in depth in \cite{pt1}
(see also \cite{pt2}), with a different approach exploiting Jordan Pairs
\cite{loos}; in the case of $\mathbf{e}_{\mathbf{8}}$, it has been also
recalled in another contribution to Group32 Proceedings \cite{3bis}.

Here, we focus on the properties of the corresponding (not necessarily
maximal, generally non-symmetric) \textit{\textquotedblleft Magic Star"
embedding}\footnote{%
We briefly recall that, in the context of supergravity, the \textit{%
\textquotedblleft Magic Star" embedding }(\ref{1}) has been discussed in
\cite{FMZ-D=5} (where it was named \textit{Jordan pairs' embedding}), and
later identified as the $D=5$ case of the so-called \textit{super-Ehlers
embedding} in \cite{super-Ehlers}. In \cite{FMZ-D=5} the \textit{%
\textquotedblleft Magic Star" embedding} for (suitable real forms of)
orthogonal Lie algebras was discussed, as well.}%
\begin{equation}
\mathbf{qconf}\left( \mathbf{J}_{3}^{q}\right) \supset \mathbf{a}_{\mathbf{2}%
}\oplus \mathbf{str}_{0}\left( \mathbf{J}_{3}^{q}\right) ,  \label{1}
\end{equation}%
and on its real forms related to a Jordan-algebraic interpretations.

Over $\mathbb{C}$, for all finite-dimensional exceptional Lie algebras (and
some classical Lie algebras) the branching corresponding to (\ref{1}) reads
as follows \cite{pt1,pt2} :%
\begin{equation}
\mathbf{qconf}\left( \mathbf{J}_{3}^{q}\right) =\mathbf{a}_{\mathbf{2}%
}\oplus \mathbf{str}_{0}\left( \mathbf{J}_{3}^{q}\right) \oplus \mathbf{3}%
\times \mathbf{J}_{3}^{q}\oplus \overline{\mathbf{3}}\times \overline{%
\mathbf{J}_{3}^{q}}.  \label{2}
\end{equation}%
By suitably varying the parameter $q$, the sequence given in Table \ref%
{grouptheorytable} is obtained. The sequence $\left\{\mathbf{qconf}\left( \mathbf{J}_{3}^{q}\right)\right\} _{q=8,4,2,1,0,-2/3,-1}$ is
usually named \textit{exceptional sequence} (or \textit{exceptional series};
cfr. e.g.\cite{LM-1}), with the addition of an $\mathbf{a}_{\mathbf{1}}$
corresponding to $q=-4/3$ (which we disregard\footnote{%
On the other hand, we add a $\mathbf{b}_{3}$ corresponding to $q=-1/3$,
which is absent in the treatment e.g. of \cite{LM-1}.}, because it is
irrelevant for our treatment).

\begin{table}[h!]
\begin{center}
\begin{tabular}{|c||c|c|c|c|c|c|c|c|}
\hline
$q$ & $8$ & $4$ & $2$ & $1$ & $0$ & $-1/3$ & $-2/3$ & $-1$ \\ \hline
\rule[-1mm]{0mm}{6mm} $\mathbf{qconf}\left( \mathbf{J}_{3}^{q}\right) $ & $\mathbf{e}_{8}$ & $\mathbf{e}_{7}$
& $\mathbf{e}_{6}$ & $\mathbf{f}_{4}$ & $\mathbf{d}_{4}$ & $\mathbf{b}_{3}$
& $\mathbf{g}_{2}$ & $\mathbf{a}_{2}$ \\ \hline
\rule[-1mm]{0mm}{6mm} $\mathbf{str}_{0}(\mathbf{J}_{3}^{q})$ & $\mathbf{e}%
_{6}$ & $\mathbf{a}_{5}$ & $\mathbf{a}_{2}\oplus \mathbf{a}_{2}$ & $\mathbf{a%
}_{2}$ & $\mathbb{C}\oplus \mathbb{C}$ & $\mathbb{C}$ & $\varnothing $ & $-$
\\ \hline
\end{tabular}%
\end{center}
\caption{The exceptional sequence }
\label{grouptheorytable}
\end{table}

$\mathbf{J}_{3}^{q}$ denotes the rank-$3$ Jordan algebra (cfr. e.g. \cite{mc}%
, and Refs. therein) associated to the parameter $q$; such a Jordan algebra
is simple for $q=8,4,2,1$ and $-2/3$ \cite{JVNW}, and for $q=8,4,2,1$ the
parameter $q$ has the meaning of real dimension of the division algebra $%
\mathbb{A}$ on which the corresponding Jordan algebra is realized as the
algebra of Hermitian matrices : $q=\dim _{\mathbb{R}}\mathbb{A}=8,4,2,1$ for
$\mathbb{A}=\mathbb{O},\mathbb{H},\mathbb{C},\mathbb{R}$, resp., and $%
\mathbf{J}_{3}^{q}\equiv \mathbf{J}_{3}^{\mathbb{A}}\equiv H_{3}\left(
\mathbb{A}\right) $ in these cases. Also, it should be remarked that for $%
q=8,4,2,1$ $\mathbf{qconf}\left( \mathbf{J}_{3}^{q}\right) $ and $\mathbf{str}_{0}\left( \mathbf{J}%
_{3}^{q}\right) $ respectively span the fourth and second row/column of the
Freudenthal-Rosenberg-Tits Magic Square \cite{tits1,freu}. Moreover\footnote{%
Note that the $q=-1$ case is a limiting case, included for completeness'
sake. In fact, in such a case there is no corresponding rank-$3$ Jordan
algebra, and so the corresponding reduced structure symmetry is ill defined.}%
, $\mathbf{J}_{3}^{0}\equiv \mathbb{C}\oplus \mathbb{C}\oplus \mathbb{C}$ is
the completely factorized (triality symmetric) rank-3 Jordan algebra, and $%
\mathbf{J}_{3}^{-1/3}\equiv \mathbb{C}\oplus \mathbb{C}$ and $\mathbf{J}%
_{3}^{-2/3}\equiv \mathbb{C}$ are its partial and total degenerations,
respectively. $\mathbf{qconf}\left( \mathbf{J}_{3}^{q}\right) $ and $\mathbf{%
str}_{0}\left( \mathbf{J}_{3}^{q}\right) $ stand for the quasi-conformal Lie
algebra resp. the reduced structure Lie algebra of the rank-3 Jordan algebra
$\mathbf{J}_{3}^{q}$ (cfr. e.g. \cite{GKN,GP-04}).

Over $\mathbb{R}$ (as we shall consider throughout the present paper), there
are two non-compact real forms of the above \textquotedblleft enlarged"
exceptional sequence $\left\{ \mathbf{qconf}\left( \mathbf{J}_{3}^{q}\right) \right\}
_{q=8,4,2,1,0,-1/3,-2/3,-1}$ which enjoy an immediate Jordan-algebraic
interpretation; they are reported in Tables \ref{grouptheorytable2} and
Table \ref{grouptheorytable3}. In both these cases, the real form of the $%
\mathbf{a}_{2}$ defining the plane onto which the Magic Star projection is
defined is maximally non-compact (i.e., split), and the following branching
(non-compact, real form of (\ref{2})) correspondingly holds:%
\begin{equation}
\mathbf{qconf}\left( \mathbf{J}_{3}^{q}\right) =\mathbf{sl}_{3,\mathbb{R}%
}\oplus \mathbf{str}_{0}\left( \mathbf{J}_{3}^{q}\right) \oplus \mathbf{3}%
\times \mathbf{J}_{3}^{q}\oplus \mathbf{3}^{\prime }\times \mathbf{J}%
_{3}^{q\prime }.  \label{2tris}
\end{equation}%
\begin{table}[h]
\begin{center}
\begin{tabular}{|c||c|c|c|c|c|c|c|c|}
\hline
$q$ & $8$ & $4$ & $2$ & $1$ & $0$ & $-1/3$ & $-2/3$ & $-1$ \\ \hline
\rule[-1mm]{0mm}{6mm} $\mathbf{qconf}\left( \mathbf{J}_{3}^{q}\right) $ & $%
\mathbf{e}_{8(8)}$ & $\mathbf{e}_{7(7)}$ & $\mathbf{e}_{6(6)}$ & $\mathbf{f}%
_{4(4)}$ & $\mathbf{so}_{4,4}$ & $\mathbf{so}_{4,3}$ & $\mathbf{g}_{2(2)}$ &
$\mathbf{sl}_{3,\mathbb{R}}$ \\ \hline
\rule[-1mm]{0mm}{6mm} $\mathbf{str}_{0}(\mathbf{J}_{3}^{q})$ & $\mathbf{e}%
_{6(6)}$ & $\mathbf{sl}_{6,\mathbb{R}}$ & $sl_{3,\mathbb{R}}\oplus sl_{3,%
\mathbb{R}}$ & $\mathbf{sl}_{3,\mathbb{R}}$ & $\mathbb{R}\oplus \mathbb{R}$
& $\mathbb{R}$ & $\varnothing $ & $-$ \\ \hline
\end{tabular}%
\end{center}
\caption{The maximally non-compact (split) real form of the exceptional
sequence. In this case, for $q=8,4,2,1$, $\mathbf{J}_{3}^{q}\equiv \mathbf{J}%
_{3}^{\mathbb{A}_{s}}\equiv H_{3}(\mathbb{A}_{s})$, where $\mathbb{A}_{s}$
is the split form of $\mathbb{A}=\mathbb{O},\mathbb{H},\mathbb{C},\mathbb{R}$%
, respectively. }
\label{grouptheorytable2}
\end{table}
\begin{table}[h]
\begin{center}
\begin{tabular}{|c||c|c|c|c|c|c|c|c|}
\hline
$q$ & $8$ & $4$ & $2$ & $1$ & $0$ & $-1/3$ & $-2/3$ & $-1$ \\ \hline
\rule[-1mm]{0mm}{6mm} $\mathbf{qconf}\left( \mathbf{J}_{3}^{q}\right) $ & $%
\mathbf{e}_{8(-24)}$ & $\mathbf{e}_{7(-5)}$ & $\mathbf{e}_{6(2)}$ & $\mathbf{%
f}_{4(4)}$ & $\mathbf{so}_{4,4}$ & $\mathbf{so}_{4,3}$ & $\mathbf{g}_{2(2)}$
& $\mathbf{sl}_{3,\mathbb{R}}$ \\ \hline
\rule[-1mm]{0mm}{6mm} $\mathbf{str}_{0}(\mathbf{J}_{3}^{q})$ & $\mathbf{e}%
_{6(-26)}$ & $\mathbf{su}_{6}^{\ast }$ & $(\mathbf{sl}_{3,\mathbb{C}})_{%
\mathbb{R}}$ & $\mathbf{sl}_{3,\mathbb{R}}$ & $\mathbb{R}\oplus \mathbb{R}$
& $\mathbb{R}$ & $\varnothing $ & $-$ \\ \hline
\end{tabular}%
\end{center}
\caption{Another (non-split) non-compact real form of the exceptional
sequence }
\label{grouptheorytable3}
\end{table}

\section{The Spin-Factor Embedding and the Exceptional Periodicity}

Let us now consider the following (maximal) algebraic embeddings ($q=8,4,2,1$%
):
\begin{eqnarray}
\mathbf{J}_{3}^{\mathbb{A}} &\supset &\mathbb{R}\oplus \mathbf{J}_{2}^{%
\mathbb{A}};  \label{A} \\
\mathbf{J}_{3}^{\mathbb{A}_{s}} &\supset &\mathbb{R}\oplus \mathbf{J}_{2}^{%
\mathbb{A}_{s}},  \label{As}
\end{eqnarray}%
realized as\footnote{%
The bar denotes the conjugation in $\mathbb{A}$ or in $\mathbb{A}_{s}$.} ($%
r_{i}\in \mathbb{R}$, $A_{i}\in \mathbb{A}$ or $\mathbb{A}_{s}$, $i=1,2,3$)%
\begin{equation}
\mathbf{J}_{3}^{\mathbb{A}}\ni J=\left(
\begin{array}{ccc}
r_{1} & A_{1} & \overline{A}_{2} \\
\overline{A}_{1} & r_{2} & A_{3} \\
A_{2} & \overline{A}_{3} & r_{3}%
\end{array}%
\right) \mapsto J^{\prime }=\left(
\begin{array}{ccc}
r_{1} & A_{1} & 0 \\
\overline{A}_{1} & r_{2} & 0 \\
0 & 0 & r_{3}%
\end{array}%
\right) \in \mathbb{R}\oplus \mathbf{J}_{2}^{\mathbb{A}}.
\end{equation}%
By noticing that $r_{1}$ and $r_{2}$ can be associated to lightcone degrees
of freedom,%
\begin{equation}
r_{1}:=x_{+}+x_{-},~r_{2}:=x_{+}-x_{-},  \label{lightcone}
\end{equation}%
it can be proved that (cfr. e.g. \cite{29-of-EYM})%
\begin{eqnarray}
\mathbf{J}_{2}^{\mathbb{A}} &\sim &\mathbf{\Gamma }_{1,q+1}; \\
\mathbf{J}_{2}^{\mathbb{A}_{s}} &\sim &\mathbf{\Gamma }_{q/2+1,q/2+1},
\end{eqnarray}%
where $\mathbf{\Gamma }_{1,q+1}$ and $\mathbf{\Gamma }_{q/2+1,q/2+1}$
respectively are $\left( q+2\right) $-dimensional spin factors with
Lorentzian and split (Kleinian) signature, respectively; thus, we will dub (%
\ref{A})-(\ref{As}) the \textit{spin-factor embeddings}.\bigskip\

Let us now restrict to consider the embedding (\ref{A}) for $\mathbb{A}=%
\mathbb{O}$, i.e. for $q=8$, and let us analyze its consequences at the
level of the symmetry algebras of its l.h.s. and r.h.s. :

\begin{enumerate}
\item at the level of the Lie algebra of derivations $\mathfrak{der}$, it
holds that\footnote{%
The upper scripts \textquotedblleft $m$" and \textquotedblleft $s$"
respectively denote the maximality and symmetricity of the embedding under
consideration.}%
\begin{equation}
\mathfrak{der}\left( \mathbf{J}_{3}^{\mathbb{O}}\right) \supset ^{m,s}%
\mathfrak{der}\left( \mathbb{R}\oplus \mathbf{J}_{2}^{\mathbb{O}}\right)
\Leftrightarrow \left\{
\begin{array}{c}
\mathbf{f}_{4(-52)}\supset ^{m,s}\mathbf{so}_{9}; \\
~ \\
\mathbf{52}=\mathbf{36}\oplus \mathbf{16},%
\end{array}%
\right.
\end{equation}%
where $\mathbf{16}$ is the real spinor irrepr. of $\mathbf{so}_{9}$.

\item at the level of the reduced structure Lie algebra $\mathfrak{str}_{0}$%
, it holds that%
\begin{equation}
\mathfrak{str}_{0}\left( \mathbf{J}_{3}^{\mathbb{O}}\right) \supset ^{m,s}%
\mathfrak{str}_{0}\left( \mathbb{R}\oplus \mathbf{J}_{2}^{\mathbb{O}}\right)
\Leftrightarrow \left\{
\begin{array}{l}
\mathbf{e}_{6(-26)}\supset ^{m,s}\mathbf{so}_{9,1}\oplus \mathbb{R}; \\
~ \\
\begin{array}{c}
\mathbf{78}=\mathbf{16}_{-1}^{\prime }\oplus \left( \mathbf{45}\oplus
\mathbf{1}\right) _{0}\oplus \mathbf{16}_{1}, \\
\text{or} \\
\mathbf{78}=\mathbf{16}_{-1}\oplus \left( \mathbf{45}\oplus \mathbf{1}%
\right) _{0}\oplus \mathbf{16}_{1}^{\prime },%
\end{array}%
\end{array}%
\right.
\end{equation}%
where $\mathbf{16}$ and $\mathbf{16}^{\prime }$ are the Majorana-Weyl (MW)
spinors of $\mathbf{so}_{9,1}$; the indeterminacy denoted by
\textquotedblleft or" depends on the \textit{spinor polarization} \cite%
{Minchenko}.

\item at the level of the conformal Lie algebra $\mathfrak{conf}$, it holds
that%
\begin{equation}
\mathfrak{conf}\left( \mathbf{J}_{3}^{\mathbb{O}}\right) \supset ^{m,s}%
\mathfrak{conf}\left( \mathbb{R}\oplus \mathbf{J}_{2}^{\mathbb{O}}\right)
\Leftrightarrow \left\{
\begin{array}{l}
\mathbf{e}_{7(-25)}\supset ^{m,s}\mathbf{so}_{10,2}\oplus \mathbf{sl}_{2,%
\mathbb{R}}; \\
~ \\
\mathbf{133}=\left( \mathbf{78,1}\right) \oplus \left( \mathbf{1,3}\right)
\oplus \left( \left( \mathbf{32}^{(\prime )},\mathbf{2}\right) \right) ,%
\end{array}%
\right.
\end{equation}%
where $\mathbf{32}$ is the MW spinor of $\mathbf{so}_{10,2}$, and the
possible priming (denoting spinorial conjugation) depends on the choice of
the spinor polarization \cite{Minchenko}. By further branching the $\mathbf{%
sl}_{2,\mathbb{R}}$, one gets the following 5-grading of contact type :%
\begin{equation}
\begin{array}{l}
\mathbf{e}_{7(-25)}\supset \mathbf{so}_{10,2}\oplus \mathbb{R}; \\
~ \\
\mathbf{133}=\mathbf{1}_{-2}\oplus \mathbf{32}_{-1}^{(\prime )}\oplus \left(
\mathbf{so}_{10,2}\oplus \mathbb{R}\right) _{0}\oplus \mathbf{32}%
_{1}^{(\prime )}\oplus \mathbf{1}_{2}.%
\end{array}%
\end{equation}

\item at the level of the quasi-conformal Lie algebra $\mathfrak{qconf}$, it
holds that%
\begin{equation}
\mathfrak{qconf}\left( \mathbf{J}_{3}^{\mathbb{O}}\right) \supset ^{m,s}%
\mathfrak{qconf}\left( \mathbb{R}\oplus \mathbf{J}_{2}^{\mathbb{O}}\right)
\Leftrightarrow \left\{
\begin{array}{l}
\mathbf{e}_{8(-24)}\supset ^{m,s}\mathbf{so}_{12,4}; \\
~ \\
\mathbf{248}=\mathbf{120}\oplus \mathbf{128}^{(\prime )},%
\end{array}%
\right.
\end{equation}%
where $\mathbf{128}$ is the MW spinor of $\mathbf{so}_{12,4}$, and again the
possible priming (denoting spinorial conjugation) depends on the choice of
the spinor polarization \cite{Minchenko}. By further branching the $\mathbf{%
so}_{12,4}$, one gets the following 5-grading of \textquotedblleft extended
Poincar\'{e}" type \cite{Cantarini-Ricciardo-Santi}:%
\begin{equation}
\begin{array}{l}
\mathbf{e}_{8(-24)}\supset \mathbf{so}_{11,3}\oplus \mathbb{R}; \\
~ \\
\mathbf{248}=\left\{
\begin{array}{c}
\mathbf{14}_{-2}\oplus \mathbf{64}_{-1}^{\prime }\oplus \left( \mathbf{so}%
_{11,3}\oplus \mathbb{R}\right) _{0}\oplus \mathbf{64}_{1}\oplus \mathbf{14}%
_{2}; \\
\text{or} \\
\mathbf{14}_{-2}\oplus \mathbf{64}_{-1}\oplus \left( \mathbf{so}%
_{11,3}\oplus \mathbb{R}\right) _{0}\oplus \mathbf{64}_{1}^{\prime }\oplus
\mathbf{14}_{2},%
\end{array}%
\right.%
\end{array}%
\end{equation}%
where $\mathbf{64}$ is the MW spinor of $\mathbf{so}_{11,3}$ and, again,
there is some indeterminacy depending on the spinor polarization \cite%
{Minchenko}.
\end{enumerate}

By exploiting Bott periodicity for the spinor irreprs., one can formally
define the \textit{\textquotedblleft Exceptional Periodicity generalizations"%
} \cite{trm1,3bis} of the real forms of exceptional Lie algebras $\mathfrak{%
der}\left( \mathbf{J}_{3}^{\mathbb{O}}\right) =\mathbf{f}_{4(-52)}$, $%
\mathfrak{str}_{0}\left( \mathbf{J}_{3}^{\mathbb{O}}\right) =\mathbf{e}%
_{6(-26)}$, $\mathfrak{conf}\left( \mathbf{J}_{3}^{\mathbb{O}}\right) =%
\mathbf{e}_{7(-25)}$ and $\mathfrak{qconf}\left( \mathbf{J}_{3}^{\mathbb{O}%
}\right) =\mathbf{e}_{8(-24)}$ (or, more briefly, the following real forms
of \textit{Exceptional Periodicity algebras}), as follows\footnote{%
In the EP generalization, for simplicity's sake, we assume to have chosen a
spinor polarization, so as to remove the \textit{a priori} indeterminacy in
the vector space grading structure of the EP algebras.} ($n\in \mathbb{N\cup
}\left\{ 0\right\} $ throughout\footnote{%
Note that the index $n$ used here is actually $n-1$, where $n$ is the index
used in \cite{3bis} and \cite{trm1}. In other words, the exceptional Lie
algebras (trivial level of EP) are obtained for $n=0$ in the present paper,
whereas they are obtained for $n=1$ in \cite{3bis} and \cite{trm1}.}) :

\begin{enumerate}
\item Exceptional periodization of level $\mathfrak{der}$ :
\begin{equation}
\mathbf{f}_{4(-52)}^{(n)}:=\mathbf{so}_{9+8n}\oplus \mathbf{\psi }_{\mathbf{%
so}_{9+8n}},  \label{EP-1}
\end{equation}%
where $\mathbf{\psi }_{\mathbf{so}_{9+8n}}\equiv \boldsymbol{2}^{4+4n}$~is
the real~spinor~of~$\mathbf{so}_{9+8n}.$

\item Exceptional periodization of level $\mathfrak{str}_{0}$ :%
\begin{equation}
\mathbf{e}_{6(-26)}^{(n)}:=\mathbf{\psi }_{\mathbf{so}_{9+8n,1},-1}^{\prime
}\oplus \left( \mathbf{so}_{9+8n,1}\oplus \mathbb{R}\right) _{0}\oplus
\mathbf{\psi }_{\mathbf{so}_{9+8n,1},1},  \label{EP-2}
\end{equation}%
where $\mathbf{\psi }_{\mathbf{so}_{9+8n,1}}\equiv \boldsymbol{2}^{4+4n}$~is
the MW~spinor~of~$\mathbf{so}_{9+8n,1}.$

\item Exceptional periodization of level $\mathfrak{conf}$ :%
\begin{equation}
\begin{array}{l}
\mathbf{e}_{7(-25)}^{(n)}:=\left( \mathbf{so}_{10+8n,2}\oplus \mathbf{sl}_{2,%
\mathbb{R}}(2,\mathbb{R})\right) \oplus \left( \mathbf{\psi }_{\mathbf{so}%
_{10+8n,2}},\mathbf{2}\right) = \\
~ \\
\mathbf{1}_{-2}\oplus \mathbf{\psi }_{\mathbf{so}_{10+8n,2},-1}\oplus \left(
\mathbf{so}_{10+8n,2}\oplus \mathbf{sl}_{2,\mathbb{R}}(2,\mathbb{R})\right)
_{0}\oplus \mathbf{\psi }_{\mathbf{so}_{10+8n,2},1}\oplus \mathbf{1}_{2},%
\end{array}
\label{EP-3}
\end{equation}%
where $\mathbf{\psi }_{\mathbf{so}_{10+8n,2}}\equiv \boldsymbol{2}^{5+4n}$%
~is the MW~spinor~of~$\mathbf{so}_{10+8n,2}.$

\item Exceptional periodization of level $\mathfrak{qconf}$ :%
\begin{equation}
\begin{array}{l}
\mathbf{e}_{8(-24)}^{(n)}:=\mathbf{so}_{12+8n,4}\oplus \mathbf{\psi }_{%
\mathbf{so}_{12+8n,4}} \\
~ \\
=\left( \mathbf{14+8n}\right) _{-2}\oplus \mathbf{\psi }_{\mathbf{so}%
_{11+8n,3},-1}^{\prime }\oplus \left( \mathbf{so}_{11+8n,3}\oplus \mathbb{R}%
\right) _{0}\oplus \mathbf{\psi }_{\mathbf{so}_{11+8n,3},1}\oplus \left(
\mathbf{14+8n}\right) _{2},%
\end{array}
\label{EP-4}
\end{equation}%
where $\mathbf{\psi }_{\mathbf{so}_{12+8n,4}}\equiv \boldsymbol{2}^{7+4n}$
and $\mathbf{\psi }_{\mathbf{so}_{11+8n,3}}\equiv \boldsymbol{2}^{6+4n}$
respectively denote the MW spinors of $\mathbf{so}_{12+8n,4}$ and of $%
\mathbf{so}_{11+8n,3}$.
\end{enumerate}

So far, (\ref{EP-1})-(\ref{EP-2}) is just a bunch of definitions, exploiting
Bott periodicity. On $\mathbb{C}$, in \cite{trm1} (see also \cite{3bis}) the
rigorous definitions of \textit{EP algebras} $\mathbf{f}_{4}^{(n)}$, $%
\mathbf{e}_{6}^{(n)}$, $\mathbf{e}_{7}^{(n)}$ and $\mathbf{e}_{8}^{(n)}$ was
dealt with by introducing the generalized roots and by defining the
commutations relations of the corresponding generators with (a suitably
generalized) Kac's asymmetry function \cite{graaf,Kac}. Here, we would like
to recall that EP algebras are not simply non-reductive, spinorial
extensions of Lie algebras, but rather they are characterized by a
non-translational (i.e., non-Abelian) nature of their spinorial sector; this
implies that they are Lie algebras only for $n=0$, i.e. at the trivial level
of EP, whereas for $n\geqslant 1$ they are not Lie algebras, because the
Jacobi identity is violated in the spinorial sector itself \cite{trm1}.

The above treatment on $\mathbb{R}$, based on the EP generalization of the
symmetry Lie algebras of $\mathbf{J}_{3}^{\mathbb{O}}$, allowed to determine
some of the real forms of the \textit{EP algebras} $\mathbf{f}_{4}^{(n)}$, $%
\mathbf{e}_{6}^{(n)}$, $\mathbf{e}_{7}^{(n)}$ and $\mathbf{e}_{8}^{(n)}$.
The other (compact and non-compact) real forms can be analogously\footnote{%
Apart from the EP-generalization $\mathbf{f}_{4(-52)}^{(n)}$ (\ref{EP-1}) of
the compact real form $\mathbf{f}_{4(-52)}$, the EP-generalized non-compact
real forms $\mathbf{e}_{6(-26)}^{(n)}$ (\ref{EP-2}), $\mathbf{e}%
_{7(-25)}^{(n)}$ (\ref{EP-3}) and $\mathbf{e}_{8(-24)}^{(n)}$ (\ref{EP-4}),
albeit being well defined, are characterized by a threefold degree of
arbitrariness. Indeed, at the $n$-th level of EP, we understood to enlarge
by $8n$ only the spacelike dimensions in the $\left( s,t\right) $-signature
of the reductive, pseudo-orthogonal part of the aforementioned EP algebras,
thus obtaining $\mathbf{so}_{9+8n,1}$, $\mathbf{so}_{10+8n,2}$ and $\mathbf{%
so}_{12+8n,4}$ (and then $\mathbf{so}_{11+8n,3}$), resp. in $\mathbf{e}%
_{6(-26)}^{(n)}$, $\mathbf{e}_{7(-25)}^{(n)}$ and $\mathbf{e}_{8(-24)}^{(n)}$%
. Nevertheless, the conjugation and reality properties of spinors depend
only on $D=s+t$ and on $\rho :=s-t$ (cfr. e.g. \cite{Spinor Algebras}), thus
at the $n$-th level of EP, the implementation of Bott (i.e., mod($8n$))
periodicity could also be made by increasing by $8n$ only the timelike
dimensions, or also by increasing by $4n$ both spacelike and timelike
dimensions. In the former case, one would obtain $\mathbf{so}_{9,1+8n}$, $%
\mathbf{so}_{10,2+8n}$ and $\mathbf{so}_{12,4+8n}$ (and then $\mathbf{so}%
_{11,3+8n}$) resp. in $\mathbf{e}_{6(-26)}^{(n)}$, $\mathbf{e}%
_{7(-25)}^{(n)} $ and $\mathbf{e}_{8(-24)}^{(n)}$, whereas in the latter
case one would obtain $\mathbf{so}_{9+4n,1+4n}$, $\mathbf{so}_{10+4n,2+4n}$
and $\mathbf{so}_{12+4n,4+4n}$ (and then $\mathbf{so}_{11+4n,3+4n}$) resp.
in $\mathbf{e}_{6(-26)}^{(n)}$, $\mathbf{e}_{7(-25)}^{(n)}$ and $\mathbf{e}%
_{8(-24)}^{(n)}$. Such a threefold degeneracy of the implementation of Bott
periodicity (yielding spinors with the same dimensions, reality properties
and conjugation properties) can in principle be applied at any level of EP,
also in a different way from the way it was implemented at the previous
level; this allows to span a large variery of $\left( s,t\right) $%
-signatures in the $\mathbf{so}_{s,t}$ reductive part of the non-compact
real forms of EP algebras.} defined e.g. by considering $\mathbf{J}_{3}^{%
\mathbb{O}_{s}}$; we plan to deal with a rigorous definition of the real
forms of EP algebras, through the generalized roots and the definition of
suitable involutions in the corresponding EP lattice \cite{trm1}, in a
forthcoming investigation.

\begin{figure}[h]
\centering
\includegraphics[width=0.60\textwidth]{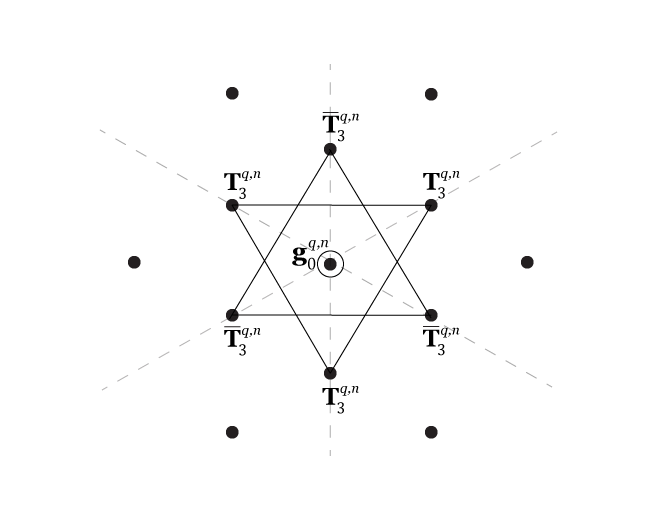}
\caption{The \textit{\textquotedblleft Magic Star"} of EP-generalized,
finite-dimensional exceptional Lie algebras \protect\cite{trm1}. $\mathbf{T}%
_{3}^{q,n}$ denotes a T-algebra of rank-3 and of \textit{special type}
\protect\cite{Vinberg}, parametrized by $q=\dim _{\mathbb{R}}\mathbb{A}%
=1,2,4,8$ for $\mathbb{A}=\mathbb{R},\mathbb{C},\mathbb{H},\mathbb{O}$, and $%
n\in N\cup \left\{ 0\right\} $, corresponding to $\mathbf{f}_{4}^{(n)}$, $%
\mathbf{e}_{6}^{(n)}$, $\mathbf{e}_{7}^{(n)}$, $\mathbf{e}_{8}^{(n)}$,
respectively. The smallest exceptional Lie algebra $\mathbf{g}_{2}$
(corresponding to $q=-2/3$) cannot be EP-generalized, because it does not
enjoy a spin factor embedding, and because $\mathbf{J}_{3}^{-2/3}\equiv
\mathbb{I}$.}
\label{fig:MagicStarT}
\end{figure}

The crucial result, which all the above construction and the corresponding
construction in the EP lattice non-trivial, is the following \cite{trm1}:
there exists an\footnote{%
Such a projection is not unique; as for the Magic Star of Lie algebras,
depicted in Figure \ref{fig:MagicStar}, also for the Magic Star of EP
algebras, depicted in Figure \ref{fig:MagicStarT}, there are four possible
equivalent projections.} $\mathbf{a}_{2}$ projection of the EP algebras
(namely, of the corresponding EP lattice), such that \textit{a Magic Star
structure persists}, with suitable generalizations of rank-$3$ Jordan
algebras $J_{3}^{n}$ (given by rank-3 T-algebras \textit{of special type}
\cite{Vinberg}) occurring on the tips of the persisting Magic Star! The
resulting, EP-generalized Magic Star is depicted in Fig. \ref{fig:MagicStarT}%
.

\section{Vinberg's T-Algebras in EP : a Glimpse}

In the treatment of previous Section, we fixed $q=8$ and dealt with the
EP-generalization of the Lie algebras which are symmetries of the
corresponding rank-3 simple Jordan algebra $\mathbf{J}_{3}^{\mathbb{O}}$.
Actually, due to the symmetry of the Freudenthal-Rosenberg-Tits Magic Square
\cite{tits1,freu} (over $\mathbb{C}$), the $\mathfrak{der}$, $\mathfrak{str}%
_{0}$, $\mathfrak{conf}$ and $\mathfrak{qconf}$ Lie symmetries (sitting in
fourth row of the Magic Square) also corresponds to the fourth column of the
Magic Square itself, and thus to $q=1,2,4,8$, respectively. Such a
correspondence, pointed out in the caption of Fig. \ref{fig:MagicStarT},
will be exploited in the treatment below.

Concerning the EP-generalized real forms treated in the previous Section,
namely $\mathbf{e}_{8(-24)}^{(n)}$, $\mathbf{e}_{7(-25)}^{(n)}$, $\mathbf{e}%
_{6(-26)}^{(n)}$ and $\mathbf{f}_{4(-52)}^{(n)}$ (corresponding to $q=8,4,2,1
$, respectively), the $3\times 3$ matrix algebras $\mathbf{T}_{3}^{q,n}$
occurring on the tips of the EP-generalized Magic Star depicted in Fig. \ref%
{fig:MagicStarT} are of the following form :%
\begin{equation}
\mathbf{T}_{3}^{q,n}:=\left(
\begin{array}{ccc}
r_{1} & \mathbf{V}_{\mathbf{so}_{q+8n}} & \mathbf{\psi }_{\mathbf{so}_{q+8n}}
\\
\overline{\mathbf{V}}_{\mathbf{so}_{q+8n}} & r_{2} & \mathbf{\psi }_{\mathbf{%
so}_{q+8n}}^{\prime } \\
\overline{\mathbf{\psi }}_{\mathbf{so}_{q+8n}} & \overline{\mathbf{\psi }%
^{\prime }}_{\mathbf{so}_{q+8n}} & r_{3}%
\end{array}%
\right) ,  \label{T}
\end{equation}%
where $\mathbf{V}_{\mathbf{so}_{q+8n}}\equiv \left( \boldsymbol{q+8n,1}%
\right) $ and $\mathbf{\psi }_{\mathbf{so}_{q+8n}}\equiv \left( \boldsymbol{2%
}^{q/2+4n-1},\mathbf{Fund}\left( \mathcal{S}_{q}\right) \right) $ are
representations of $\mathbf{so}_{q+8n}\oplus \mathcal{S}_{q}$, with
\begin{equation}
\mathcal{S}_{q}:=\mathbf{tri}_{\mathbb{A}}\ominus \mathbf{so}_{\mathbb{A}%
}=\varnothing ,\mathbf{so}_{2},\mathbf{su}_{2},\varnothing ~\text{for~}%
q=1,2,4,8~\text{(i.e.,~for~}\mathbb{R}\text{,}\mathbb{C}\text{,}\mathbb{H}%
\text{,}\mathbb{O}\text{,~resp.)}
\end{equation}%
being the coset algebra of the triality symmetry $\mathbf{tri}_{\mathbb{A}}$
of $\mathbb{A}$ \cite{Barton-Sudbery}:%
\begin{equation*}
\begin{array}{lll}
\mathbf{tri}_{\mathbb{A}}: & = & \left\{ \left( A,B,C\right)
|A(xy)=B(x)y+xC(y),~A,B,C\in \mathbf{so}_{\mathbb{A}},~x,y\in \mathbb{A}%
\right\}  \\
~ & ~ & ~ \\
~ & = & \varnothing ,\mathbf{so}_{2}^{\oplus 2},\mathbf{so}_{3}{}^{\oplus 3},%
\mathbf{so}_{8}~\text{for~}\mathbb{A}=\mathbb{R},\mathbb{C},\mathbb{H},%
\mathbb{O}%
\end{array}%
\end{equation*}%
modded by the norm preserving symmetry $\mathbf{so}_{\mathbb{A}}$of $\mathbb{%
A}$ :%
\begin{equation}
\mathbf{so}_{\mathbb{A}}:=\mathbf{so}_{q}=\varnothing ,\mathbf{so}_{2},%
\mathbf{so}_{4},\mathbf{so}_{8}~\text{for~}\mathbb{A}=\mathbb{R},\mathbb{C},%
\mathbb{H},\mathbb{O}.
\end{equation}%
$\mathbf{Fund}\left( \mathcal{S}_{q}\right) $ denotes the smallest
non-trivial representation of the simple Lie algebra $\mathcal{S}_{q}$ (if
any) :
\begin{equation}
\mathbf{Fund}\left( \mathcal{S}_{q}\right) =\mathbf{-},\mathbf{2},\mathbf{2}%
,-\mathbf{~}\text{for~}q=1,2,4,8,
\end{equation}%
with real dimension%
\begin{equation}
fund_{q}:=\dim _{\mathbb{R}}\mathbf{Fund}\left( \mathcal{S}_{q}\right)
=1,2,2,1\mathbf{~}\text{for~}q=1,2,4,8.
\end{equation}%
Thus, the total real dimension of $\mathbf{T}_{3}^{q,n}$ is%
\begin{equation}
\dim _{\mathbb{R}}(\mathbf{T}_{3}^{q,n})=q+3+8n+fund_{q}2^{\left[ q/2\right]
+4n+\delta _{q,1}},
\end{equation}%
where the square brackets denote the integer part.

$\mathbf{T}_{3}^{q,n}$ (\ref{T}) is a rank-3 T-algebra, introduced by
Vinberg as a generalization of cubic Jordan algebras in the study of
homogeneous convex cones \cite{Vinberg}. In particular, $\mathbf{T}_{3}^{q,n}
$ is a rank-3 T-algebra of \textit{special} type; it is dubbed \textit{%
exceptional }T-algebra in Sec. 4.3 of \cite{3bis}. Such algebras, slightly
generalized, also occur in the study of homogeneous non-symmetric real
special manifolds\footnote{%
And, of course, in their images under R-map and c-map (cfr. e.g. \cite{dWVVP}%
, and Refs. therein).}, which are non-compact Riemannian spaces occurring as
(vector multiplets') scalar manifolds of $\mathcal{N}=2$ Maxwell-Einstein
supergravity theories in $D=s+t=4+1$ space-time dimensions \cite{Cecotti}.

It is immediate to check that \textit{triality} is broken in $\mathbf{T}%
_{3}^{q,n}$ whenever $n$ is positive; in fact, triality holds whenever%
\begin{equation}
\begin{array}{c}
\dim _{\mathbb{R}}\mathbf{V}=\dim _{\mathbb{R}}\mathbf{\psi } \\
\Updownarrow  \\
q+8n=fund_{q}\cdot 2^{\left[ q/2\right] +4n-1+\delta _{q,1}} \\
\Updownarrow  \\
n=0:\mathbf{T}_{3}^{q,0}\equiv \mathbf{J}_{3}^{q}.%
\end{array}%
\end{equation}%
Interestingly, a kind of \textit{weaker} (or \textit{generalized)} form of
triality exists for\footnote{%
This was communicated to us by Eric Weinstein during an informal discussion
in 2016.} $\mathbf{so}_{24}$ : this corresponds to $q=8$ and $n=2$, i.e. to
the second non-trivial level of EP over $\mathbb{O}$. In this case, $%
\mathcal{S}_{8}=\varnothing $, and from definition (\ref{T}) $\mathbf{T}%
_{3}^{8,2}$ reads as follows :%
\begin{equation}
\mathbf{T}_{3}^{8,2}=\left(
\begin{array}{ccc}
r_{1} & \mathbf{V}_{\mathbf{so}_{24}} & \mathbf{\psi }_{\mathbf{so}_{24}} \\
\overline{\mathbf{V}}_{\mathbf{so}_{24}} & r_{2} & \mathbf{\psi }_{\mathbf{so%
}_{24}}^{\prime } \\
\overline{\mathbf{\psi }}_{\mathbf{so}_{24}} & \overline{\mathbf{\psi }%
^{\prime }}_{\mathbf{so}_{24}} & r_{3}%
\end{array}%
\right) =\left(
\begin{array}{ccc}
\mathbf{1} & \mathbf{24} & \mathbf{2048} \\
\mathbf{24} & \mathbf{1} & \overline{\mathbf{2048}} \\
\mathbf{2048} & \overline{\mathbf{2048}} & \mathbf{1}%
\end{array}%
\right) ~\text{of~}\mathbf{so}_{24}.
\end{equation}%
Remarkably, there exists an irreducible representation of $\mathbf{so}_{24}$
whose dimension is exactly the difference between the dimensions of the
semispinor ($\mathbf{2048}$) and of the vector ($\mathbf{24}$) of $\mathbf{so%
}_{24}$ : this is the 3-form representation $\wedge ^{3}\mathbf{24}=\mathbf{%
2024}$. Thus, one could argue the existence of an \textquotedblleft
augmented" T-algebra%
\begin{equation}
\widetilde{\mathbf{T}}_{3}^{8,2}:=\left(
\begin{array}{ccc}
r_{1} & \mathbf{V}_{\mathbf{so}_{24}}\oplus \wedge _{\mathbf{so}_{24}}^{3} &
\mathbf{\psi }_{\mathbf{so}_{24}} \\
\overline{\mathbf{V}}_{\mathbf{so}_{24}}\oplus \overline{\wedge ^{3}}_{%
\mathbf{so}_{24}} & r_{2} & \mathbf{\psi }_{\mathbf{so}_{24}}^{\prime } \\
\overline{\mathbf{\psi }}_{\mathbf{so}_{24}} & \overline{\mathbf{\psi }%
^{\prime }}_{\mathbf{so}_{24}} & r_{3}%
\end{array}%
\right) =\left(
\begin{array}{ccc}
\mathbf{1} & \mathbf{24\oplus 2024} & \mathbf{2048} \\
\mathbf{24\oplus 2024} & \mathbf{1} & \overline{\mathbf{2048}} \\
\mathbf{2048} & \overline{\mathbf{2048}} & \mathbf{1}%
\end{array}%
\right) ~\text{of~}\mathbf{so}_{24},  \label{T-aug}
\end{equation}%
exhibiting a manifest triality, albeit with a reducible nature of the
bosonic representations. It would be interesting to investigate the
properties of \textquotedblleft augmented" T-algebras of the type (\ref%
{T-aug}). However, it is here worth stressing that triality is rigorously
holding only for the trivial level of EP, i.e. for $n=0$. For instance, when
$n=0$ and $q=8$, one obtains the well known case of so(8) triality among the
off-diagonal blocks of the Albert algebra $\mathbf{J}_{3}^{8}\equiv \mathbf{J%
}_{3}^{\mathbb{O}}$ ($o_{i}\in \mathbb{O}$, $r_{i}\in \mathbb{R}$):%
\begin{equation}
\mathbf{T}_{3}^{8,0}\equiv \mathbf{J}_{3}^{\mathbb{O}}\ni \left(
\begin{array}{ccc}
r_{1} & o_{1} & \overline{o}_{2} \\
\overline{o}_{1} & r_{2} & o_{3} \\
o_{2} & \overline{o}_{3} & r_{3}%
\end{array}%
\right) =\left(
\begin{array}{ccc}
\mathbf{1} & \mathbf{8}_{v} & \mathbf{8}_{s} \\
\mathbf{8}_{v} & \mathbf{1} & \mathbf{8}_{c} \\
\mathbf{8}_{s} & \mathbf{8}_{c} & \mathbf{1}%
\end{array}%
\right) ,
\end{equation}%
where $\mathbf{8}_{v}$, $\mathbf{8}_{s}$ and $\mathbf{8}_{c}$ are the
vector, semispinor and conjugate semispinor of $\mathbf{tri}_{\mathbb{O}}=%
\mathbf{so}_{8}$, respectively.

For what concerns the invariant structures of the special class of rank-3
T-algebras defined in (\ref{T}), let us define ($\mu =0,1,...,q+1+8n$)%
\begin{equation*}
V^{\mu }:=\left( r_{1},r_{2},\mathbf{V}_{\mathbf{so}_{q+8n}}\right) ,
\end{equation*}%
which, by recalling (\ref{lightcone}), is recognized to be a vector in a $%
\left( q+2+8n\right) $-dimensional space, with Lorentzian signature $\left(
s,t\right) =\left( q+1+8n,1\right) $; also, let us denote the corresponding
spinor of $\mathbf{so}_{q+1+8n,1}$ (which is chiral for $q=2,4,8$), of real
dimension $fund_{q}\cdot 2^{\left[ q/2\right] +4n+\delta _{q,1}}$, by $\Psi
^{\alpha A}$ (where $\alpha =1,...,2^{\left[ q/2\right] +4n+\delta _{q,1}}$
and $A=1,fund_{q}$). Then, an invariant structure constructed with the
corresponding T-algebra $\mathbf{T}_{3}^{q,n}$ is formally given by the
\textquotedblleft determinant" of such a $3\times 3$ Hermitian matrix,
defining the cubic norm $\mathbf{N}$ of $\mathbf{T}_{3}^{q,n}$ itself :
\begin{equation}
\mathbf{N}\left( \mathbf{T}_{3}^{q,n}\right) :=\frac{1}{2}\eta _{\mu \nu }%
\left[ r_{3}V^{\mu }V^{\nu }+\gamma _{\alpha \beta }^{\mu }\Psi ^{\alpha
A}\Psi _{A}^{\beta }V^{\nu }\right] ,
\end{equation}%
where $\eta _{\mu \nu }$ is the symmetric bilinear invariant of the $\mathbf{%
q+2+8n}$ irrepr. $V$ of $\mathbf{so}_{q+1+8n,1}$ and $\gamma _{\alpha \beta
}^{\mu }$ are the gamma matrices of $\mathbf{so}_{q+1+8n,1}$. Consequently,
as for the Jordan algebras, one can classify the elements of $\mathbf{T}%
_{3}^{q,n}$ depending on their \textit{rank}, defined as follows \cite{trm1}%
:
\begin{equation}
\begin{array}{lll}
\text{rank-}3 & : & \mathbf{N}\neq 0; \\
\text{rank-}2 & : & \mathbf{N}=0; \\
\text{rank-}1 & : & \partial \mathbf{N}=0.%
\end{array}%
\end{equation}%
In the $\mathcal{N}=2$ Maxwell-Einstein supergravity theories in $D=s+t=4+1$
space-time dimensions based on $\mathbf{T}_{3}^{q,n}$ \cite{Cecotti}, the
Bekenstein-Hawking entropy $S_{BH}$ of the extremal black holes simply reads
\begin{equation}
S_{BH}=\pi \sqrt{\left\vert \mathbf{N}\right\vert }.
\end{equation}


\section*{References}


\end{document}